\documentclass[aps,prb,showpacs,twocolumn,superscriptaddress]{revtex4-2}

\usepackage{graphicx} 
\usepackage{color}
\usepackage{array}
\usepackage{mathrsfs}
\usepackage{ulem} 
\usepackage{amsmath}

\begin{document}
\title{Stability of the 1/3 magnetization plateau of the $J_1-J_2$ kagome Heisenberg model}

\author{Katsuhiro Morita}
\email[e-mail:]{katsuhiro.morita@rs.tus.ac.jp}
\affiliation{Department of Physics and Astronomy, Faculty of Science and Technology, Tokyo University of Science, Chiba 278-8510, Japan}

\begin{abstract}
 In this study, we investigate the finite-temperature properties of the spin-1/2 $J_1-J_2$ Heisenberg model on the kagome lattice using the orthogonalized finite-temperature Lanczos method.
 Under a zero magnetic field, the specific heat exhibits a double-peak structure, as $|J_2|$ increases. 
 Additionally, at approximately $J_2=0$, the magnetic entropy remains finite, even at low temperatures.
The finite-temperature magnetization curve reveals the asymmetric melting behavior of the 1/3 plateau around $J_2=0$. As $|J_2|$ increases, the 1/3 plateau becomes more stable, exhibiting symmetric melting behavior. 
Specifically, for $J_2 > 0$, the $\bf Q=0$ up-up-down structure is stabilized, whereas for $J_2 < 0$, the $\sqrt{3} \times \sqrt{3}$ up-up-down structure is stabilized.
\end{abstract}

\maketitle
\section{Introduction}
The study of the spin-1/2 kagome Heisenberg model has garnered significant attention in the fields of condensed matter physics and materials science owing to the prevalence of novel quantum phenomena \cite{KLP1,KLP2,KLP3}.
The ground state of this model is expected to be either a gapped quantum spin liquid (QSL)~\cite{Z2-1,Z2-2}, a gapless QSL~\cite{U1-1,U1-2,U1-3,U1-4}, or a valence bond crystal (VBC) state~\cite{VBC-1,VBC-2,VBC-3} because of the synergy between frustration and quantum fluctuations.
In a magnetic field at $T=0$, field-induced quantum phase transitions occur~\cite{KH1,KH2,KH3,KH4,KH5,KH6}.
The magnetization curve of this model exhibits multiple plateaus at $M/M_{\rm sat}$ = 0, 1/9, 1/3, 5/9, and 7/9, 
where $M$ is the magnetization, and $M_{\rm sat}$ is the saturation magnetization \cite{KH3, KH4, KH5}. 

Even at finite temperatures, this model has been extensively studied in recent years.
The specific heat is predicted to exhibit a distinctive multipeak structure~\cite{FT1,FT2,FT3}. 
The high-temperature peak is attributed to a crossover from the paramagnetic state to a short-range-ordered state.
Notably, the finite-temperature magnetization curve exhibits an asymmetric melting behavior of the 1/3 plateau \cite{FT3,FT4,FT5}. 
This phenomenon arises because of the significantly higher density of low-energy states in the regime with $M/M_{\rm  sat} < 1/3$ compared with the density of low-energy states in the regime with $M/M_{\rm sat} > 1/3$.
These features do not appear in the triangular lattice model, which suggests that they are caused by the strong frustration effect of the kagome lattice~\cite{FT4}.

 Several model compounds exist for $S=1/2$ kagome antiferromagnets \cite{ke1,ke2,ke3,ke4,ke5,ke6,ke7,ke8,ke9,ke10,ke11,ke12,ke13}.
In the model compounds, the next-nearest-neighbor exchange interaction $J_2$ always exists regardless of whether it is large or small.
The ground state of the $J_1-J_2$ kagome lattice has been studied theoretically \cite{J12k1,J12k2,J12k3,J12k4,J12k5}.
 For $J_1 > 0$ and  $J_2 > 0$ (where positive indicates antiferromagnetic), the ground state exhibits a $\bf Q=0$, $120^\circ$ structure, 
whereas for $J_2 < 0$, the ground state exhibits a $\sqrt{3} \times \sqrt{3}$ $120^\circ$ structure. 
In addition, it is predicted that the ground state will undergo QSL or VBC  in $-0.1\lesssim J_2/J_1 \lesssim 0.1$ \cite{J12k2,J12k3,J12k4,J12k5}.
However, theoretical studies on the kagome model with $J_2$ have primarily focused on the case of $T=0$, and there have been few studies on its finite-temperature properties.

 In this study, we investigate the finite-temperature properties of the $J_1-J_2$ kagome Heisenberg model using the orthogonalized finite-temperature Lanczos method (OFTLM) \cite{OFTL1,OFTL2}. 
In this paper, $J_1$ is set to 1 as the energy unit.
At $J_2 = 0$, the specific heat exhibits a multipeak structure; however, as $|J_2|$ increases, 
the multipeak structure transitions into a double-peak structure.
Furthermore, the position of the low-temperature peak among the two peaks gradually shifts to the high-temperature side as $|J_2|$ increases.
For $-0.05 < J_2 < 0.01$, the magnetic entropy exhibits a finite value even at low temperatures, indicating the presence of a QSL.
In the magnetization curve, similar to previous studies, we observe the asymmetric melting of the 1/3 plateau at $J_2 = 0$. In contrast, as $|J_2|$ increases, an apparent flattening of the 1/3 plateau is observed, even at $T=0.1$, and the 1/3 plateau melts symmetrically as the temperature increases.
This is because, in the region of a larger $|J_2|$, a semiclassical ground state characterized by up-up-down (uud) structures (see Fig.~\ref{1-3Sq}) similar to the structure of the 1/3 plateau in the triangular lattice emerges.
For $J_2 > 0$, the $\bf Q=0$ uud state is stable, whereas for $J_2 < 0$, the $\sqrt{3} \times \sqrt{3}$ uud state is stable. 
The region around $J_2=0$ is situated in the intermediate region between these phases, leading to the instability of the 1/3 plateau.
Consequently, the 1/3 plateau undergoes rapid melting, as the temperature increases.
Contrary to the conventional understanding that as the degeneracy of the classical ground state increases, quantum effects become more pronounced, leading to the emergence of a magnetization plateau, our findings demonstrate that the 1/3 plateau stabilizes as the degeneracy is reduced.

By comparing our results with the experimental results, we will be able to determine the value of $J_2$ for  spin-1/2 $J_1-J_2$ kagome compounds with antiferromagnetic $J_1$.
If a 1/3 plateau is not observed in the experiment, this suggests that the kagome compounds possess a relatively small $|J_2|$.

\section{Model}
\label{sec2}

\begin{figure}[tb]
\includegraphics[width=86mm]{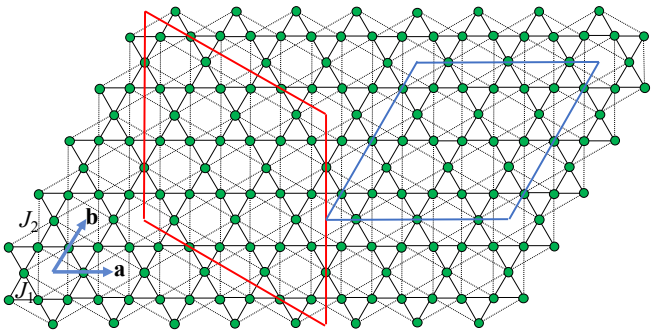}
\caption{
Lattice structure of the $J_1-J_2$ kagome lattice. 
The solid and dashed lines represent $J_1$ and $J_2$, respectively. 
$J_1$ is set to 1. 
The green circles represent the sites with a spin. 
{\bf a} and {\bf b} represent the primitive vectors of magnitude 1 ($|{\bf a}|=|{\bf b}|=1$).
The red, and blue quadrangles represent the clusters of $N = 36$ and $N = 27$ with periodic boundary conditions, respectively, where $N$ is the number of sites.
\label{LT}}
\end{figure}

The Hamiltonian for the spin-1/2 $J_1-J_2$ kagome lattice shown in Fig.~\ref{LT} in a magnetic field is defined as follows:
\begin{eqnarray} 
\mathcal{H} &=& J_{1}\sum_{\langle i,j \rangle }\mathbf{S}_i \cdot \mathbf{S}_j + J_{2}\sum_{\langle\langle i,j \rangle\rangle }\mathbf{S}_i \cdot \mathbf{S}_j - h\sum_i S^{z}_i,
\label{Hami}
\end{eqnarray}
where $\mathbf{S}_i$ is the spin-1/2 operator at the $i$-th site, $S^z_i$ is the $z$ component of $\mathbf{S}_i$, $\langle i,j \rangle$ and  $\langle\langle i,j \rangle\rangle$ run over the nearest-neighbor and next-nearest-neighbor spin pairs of the kagome lattice, respectively,
and $h$ represents the magnitude of the magnetic field applied in the $z$ direction.
Here, $J_1$ is set to 1 as the energy unit. 
In this study, we consider both positive and negative values of $J_2$.
In Heisenberg models, the operator $\sum_i S^{z}_i$ is a conserved quantity.
Here, the eigenvalue of operator $\sum_i S^{z}_i$ is defined as $S^z_{tot}$.
The vectors {\bf a} and {\bf b} shown in Fig.~\ref{LT} represent primitive vectors of magnitude one.
Main calculations are performed on a cluster consisting of 36 sites with periodic boundary conditions (PBC)  shown in Fig.~\ref{LT}.
Calculations for a cluster consisting of 27 sites with PBC are performed only to check for finite size effects.

\section{Method}
\label{sec3}
The finite-temperature Lanczos method (FTLM) has been employed in studies on frustrated quantum lattice models because it does not have the sign problem \cite{J12k4,ftla1,ftla2,ftla3,ftla4,ftla5,ftla6,ftla7,ftla8,ftla9,ftla10,ftla11,ftla12,ftla13}, which is a concern in quantum Monte Carlo simulations.
The OFTLM is a more accurate method than the standard FTLM, particularly at low temperatures.
Here, we provide a brief summary of the OFTLM ~\cite{OFTL2}.

The partition function using the standard FTLM is as follows:
\begin{equation}
\begin{split}
 Z(T,h)_{\rm FTL} = \sum_{m=-M_{\rm sat}}^{M_{\rm sat}} &\frac{N_{st}^{(m)}}{R}\sum _{r=1}^{R} \sum _{j=0}^{M_L-1}  \\
& e^{-\beta \epsilon^{(r)}_{j,m}(h)} |\langle V_{r,m} | \psi^r_{j,m} \rangle|^2, \label{ZFTL} 
\end{split}
\end{equation}
where $R$ denotes the number of random samplings of the FTLM, $M_L$ denotes the dimension of the Krylov subspace,
$|V_{r,m}\rangle$ is a normalized random initial vector with $S^z_{tot}=m$, and $|\psi^r_{j,m}\rangle$ [$\epsilon^{(r)}_{j,m}(h)$] are the eigenvectors (eigenvalues) in the $M_L$-th Krylov subspace with $S^z_{tot}=m$.
As $\sum_i S^{z}_i$ is a conserved quantity,
$\epsilon^{(r)}_{j,m}(h)$ can be expressed as $\epsilon^{(r)}_{j,m}(h)  = \epsilon^{(r)}_{j,m}  - mh$.
We define the order of \{$\epsilon^{(r)}_{j,m}$\} as $\epsilon^{(r)}_{0,m} \le \epsilon^{(r)}_{1,m}  \le \epsilon^{(r)}_{2,m} \le \cdots \le \epsilon^{(r)}_{M_L-1,m} $. 
If $M_L$ is sufficiently large, $\epsilon^{(r)}_{0,m}$ becomes equal to the exact ground state energy $E_{0,m}$.
However, $|\langle V_{r,m} | \psi^r_{j,m} \rangle|^2$ does not converge to the expected value, that is, $d_m/N_{st}^{(m)}$, where $d_m$ represents the degeneracy of the ground state in the subspace with $S^z_{tot}=m$.
Therefore, unless a sufficient number of random samples are considered, the accuracy of $Z(T,h)_{\rm FTL}$ will not improve at low temperatures.

\begin{table}[tb]
\caption{Conditions for the calculation at $N=36$}
  \begin{tabular}{ c c c c c c } \hline 
    $m$  &  $N_{st}^{(m)}$  & method & $R$ & $M_L$ & $ N_V $ \\ \hline 
   18  & 1 & Exact & -- & -- & -- \\ 
   17  & 36 & FullED & -- & -- & -- \\ 
   16  & 630 & FullED & -- & -- & -- \\ 
   15  & 7140 & FullED & -- & -- & -- \\ 
   14  & 58905 &FullED & -- & -- & -- \\ 
   13  & 376992 & OFTLM & 10 & 160 & 5   \\ 
   12  & 1947792 & OFTLM & 10 & 160 & 5   \\ 
   11  & 8347680 & OFTLM & 10 & 160 & 5   \\ 
   10  & 30260340 & OFTLM  & 10 & 160 & 5   \\ 
   9  & 94143280 & OFTLM  & 10 & 160 & 5   \\ 
   8  & 254186856 & OFTLM  & 10 & 160 & 5   \\ 
   7  & 600805296 & OFTLM  & 10 & 160 & 5   \\ 
   6  & 1251677700 & OFTLM  & 10 & 160 & 5   \\ 
   5  & 2310789600 & OFTLM  & 10 & 160 & 5   \\ 
   4  & 3796297200 & OFTLM  & 10 & 160 & 5   \\ 
   3  & 5567902560 & OFTLM  & 10 & 160 & 5   \\ 
   2  & 7307872110 & OFTLM  & 10 & 160 & 5   \\  
   1  & 8597496600 & OFTLM  & 10 & 160 & 5   \\ 
   0  & 9075135300 & OFTLM  & 10 & 160 & 5   \\ \hline 
  \end{tabular}
\label{para}
\end{table}

In the OFTLM, we first calculate several low-lying exact eigenvectors $| \Psi_{i,m} \rangle$ with $N_V$ levels.
We define the order \{$E_{i,m}$\} as $E_{0,m} \le E_{1,m}  \le \cdots \le E_{N_V-1,m} $.
We then calculate the following modulated random vector:
\begin{eqnarray} 
 |V_{r,m}'\rangle  &=& \left[ I - \sum_{i=0}^{N_V-1} | \Psi_{i,m} \rangle \langle \Psi_{i,m} |  \right] | V_{r,m} \rangle,  \label{r'}
\end{eqnarray}
with normalization
\begin{equation}
 |V_{r,m}'\rangle  \Rightarrow \frac{ |V_{r,m}'\rangle }{ \sqrt{\langle V_{r,m}' |V_{r,m}'\rangle} }. \label{r'2}
\end{equation}
The partition function of the OFTLM is obtained using $|V_{r,m}'\rangle $ as the initial vector, as follows:
\begin{equation}
\begin{split}
  Z(T,h)_{\rm OFTL} 
 &= \sum_{m=-M_{\rm sat}}^{M_{\rm sat}} \left[ \frac{N_{st}^{(m)}-N_V}{R}\sum_{r=1}^{R} \sum_{j=0}^{M_L-1}  \right. \\
& \left. e^{-\beta \epsilon^{(r)}_{j,m}(h)} |\langle V_{r,m}' | \psi^{r}_{j,m} \rangle|^2  + \sum_{i=0}^{N_V-1}  e^{-\beta E_{i,m}(h)} \right]. \label{ZOFTL} 
\end{split}
\end{equation}
If $N_V \ge d_m$, $Z(T,h)_{\rm OFTL}$ reaches its exact value at low temperatures.
Therefore, it is recommended that $N_V$ be greater than or equal to $d_m$.
Similarly, in the OFTLM, the energy $E(T)_{\rm OFTL}$, magnetic specific heat $C(T)_{\rm OFTL}$, magnetic entropy $S_{\rm m}(T)_{\rm OFTL}$ at $h=0$,
and magnetization $M(T,h)_{\rm OFTL}$ are obtained as follows:
\begin{equation}
\begin{split}
  E(T)_{\rm OFTL}  &= \frac{1}{Z(T,0)_{\rm OFTL}}\sum_{m=-M_{\rm sat}}^{M_{\rm sat}} \left[ \frac{N_{st}^{(m)}-N_V}{R}  \right. \\
      &\times \sum_{r=1}^{R} \sum _{j=0}^{M_L-1} \epsilon^{(r)}_{j,m} e^{-\beta \epsilon^{(r)}_{j,m}} |\langle V_{r,m}' | \psi^{r}_{j,m} \rangle|^2  \\
       &+ \left. \sum_{i=0}^{N_V-1} E_{i,m} e^{-\beta E_{i,m}} \right], 
\label{EOFTL} 
\end{split}
\end{equation}

\begin{equation}
\begin{split}
  C(T)_{\rm OFTL}  &= \frac{1}{T^2Z(T,0)_{\rm OFTL}}\sum_{m=-M_{\rm sat}}^{M_{\rm sat}} \left[ \frac{N_{st}^{(m)}-N_V}{R} \right.  \\
                             &\times \sum _{r=1}^{R} \sum _{j=0}^{M_L-1} {\epsilon^{(r)}_{j,m}}^2 e^{-\beta \epsilon^{(r)}_{j,m}} |\langle V_{r,m}' | \psi^{r}_{j,m} \rangle|^2  \\
                             &+ \left. \sum_{i=0}^{N_V-1} E_{i,m}^2 e^{-\beta E_{i,m}} \right] -\frac{ E(T)_{\rm OFTL}^2 }{T^2} , 
\label{COFTL} 
\end{split}
\end{equation}
\begin{equation} 
  S_{\rm m}(T)_{\rm OFTL}  = \frac{E(T)_{\rm OFTL} }{T} -\ln Z(T,0)_{\rm OFTL}.
\label{SOFTL} 
\end{equation}

\begin{equation}
\begin{split}
  M(T,h)_{\rm OFTL}  &= \frac{1}{Z(T,h)_{\rm OFTL}}\sum_{m=-M_{\rm sat}}^{M_{\rm sat}} \left[ \frac{N_{st}^{(m)}-N_V}{R}  \right.  \\
                             &\times \sum _{r=1}^{R} \sum _{j=0}^{M_L-1} m e^{-\beta \epsilon^{(r)}_{j,m}(h)} |\langle V_{r,m}' | \psi^{r}_{j,m} \rangle|^2  \\ 
                             &+ \left.  \sum_{i=0}^{N_V-1} m e^{-\beta E_{i,m}(h)} \right], 
\label{MOFTL} 
\end{split}
\end{equation}

Since Eqs.~(\ref{ZOFTL}), (\ref{EOFTL}), (\ref{COFTL}), (\ref{SOFTL}), and (\ref{MOFTL}) include the exact values $E_{i,m}$, they are more accurate than those obtained using
the standard FTLM, particularly at low temperatures.
The conditions for the calculation of the OFTLM are listed in Table~\ref{para}.
Here, FullED in Table~\ref{para} represents the full exact diagonalization method.
Note that it is possible to set values of $R$, $M_L$, and $N_V$ depending on $m$ in the OFTLM, but
we maintain them constant in the present study.
In the $J_1-J_2$ kagome lattice, when $J_2 < 0$, the ground state exhibits a $\sqrt{3} \times \sqrt{3}$ structure. Therefore, it is desirable for the lattice size $N$ to be a multiple of 9. Accordingly, in this study, we mainly perform calculations for the $N = 36$ cluster.

\begin{figure}[tb]
\includegraphics[width=86mm]{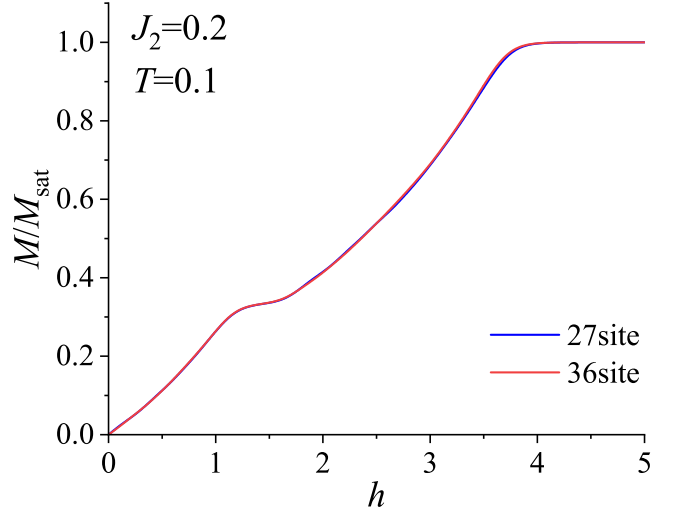}
\caption{
Magnetization curve of the $J_1-J_2$ kagome lattice at $T=0.1$ for $J_2=0.2$ with $N=27$ (blue solid line)  and $N=36$  (red solid line).
This calculation was performed to confirm the finite size effect.
\label{FSE}}
\end{figure}

\begin{figure}[tb]
\includegraphics[width=86mm]{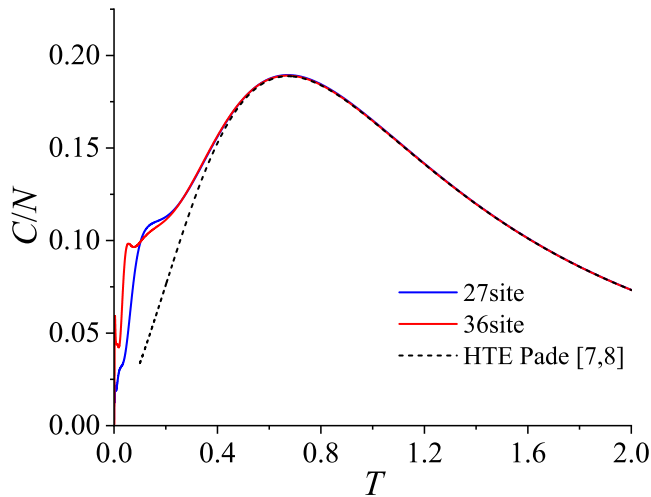}
\caption{
Specific heat of the kagome lattice ($J_2=0$) with $N=27$ (blue solid line) and $N=36$ (red solid line).
The result of the high-temperature series expansion combined with the [7,8] Pad\'{e}
approximant~\cite{FT1} is also included for comparison (black dashed line).
\label{HTE}}
\end{figure}

Before delving into the main part of the calculations, we confirm the presence of finite size effects. Figure~\ref{FSE} presents a comparison of the magnetization curves for $N=27$ and $N=36$. The two curves closely coincide, indicating that there are almost no finite-size effects for $T\ge0.1$.
Figure~\ref{HTE} presents the calculation results of the specific heat. 
The results of the 27-site and 36-site clusters exhibit a remarkable agreement for $T\gtrsim0.1$. 
Furthermore, for $T>0.4$, our results are in excellent alignment with the result of the high-temperature series expansion combined with the [7,8] Pad\'{e} approximant~\cite{FT1}.
From these results, it is evident that our calculations for the 36-site cluster accurately determine physical quantities for $T\gtrsim0.1$ in the thermodynamic limit.

\section{Results}
\label{sec4}
\subsection{Specific heat and entropy}

\begin{figure}[tb]
\includegraphics[width=86mm]{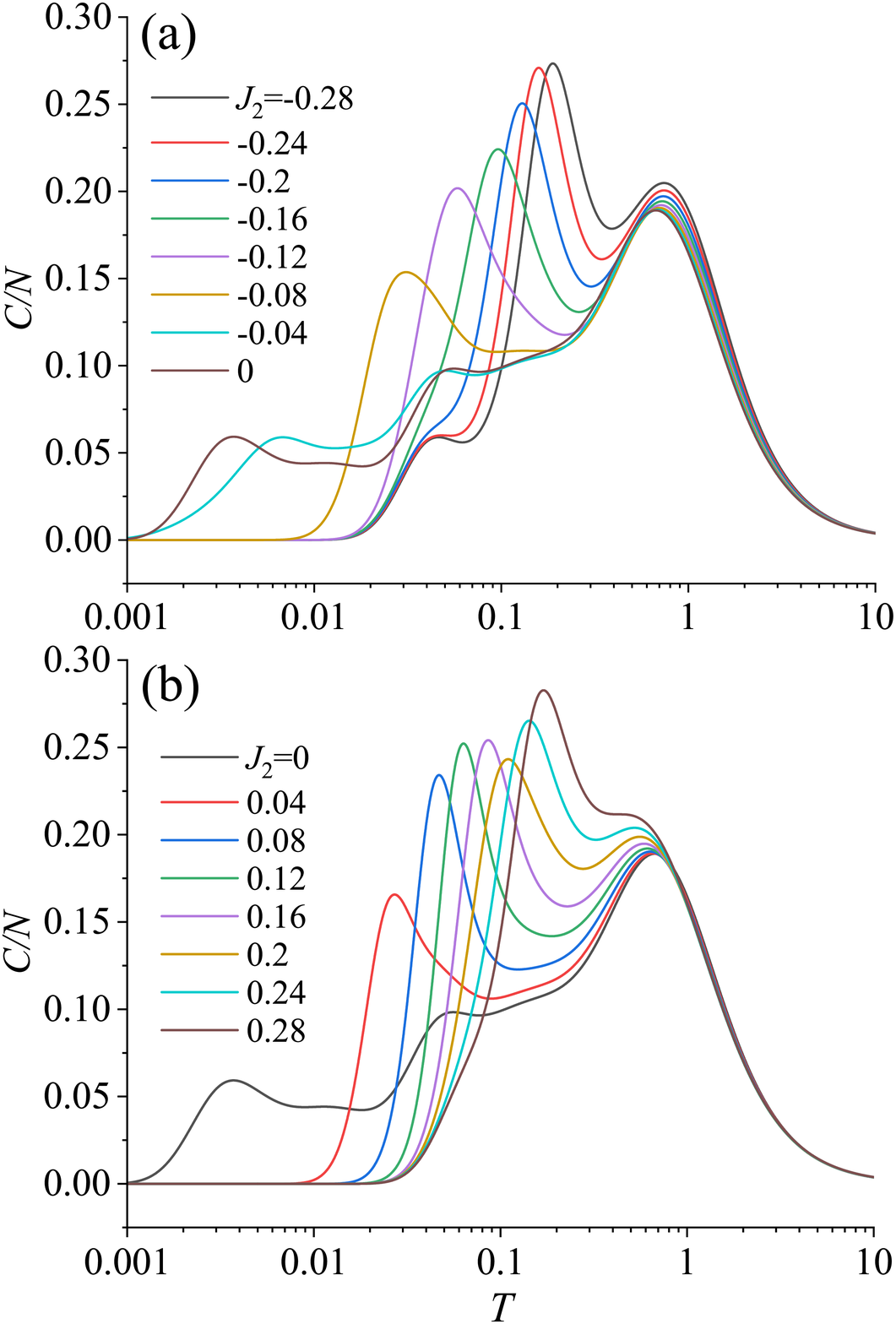}
\caption{
Temperature dependence of the specific heat per site for the $J_1-J_2$ kagome lattice with $N=36$ at $h=0$ for $J_2 \ge 0$ (a) and $J_2 \le 0$ (b).
\label{Cm}}
\end{figure}

Figure~\ref{Cm}(a) and \ref{Cm}(b) show the specific heat results for $-0.28 \leq J_2 \leq 0$ and $0 \leq J_2 \leq 0.28$, respectively.
At $J_2=0$, 
the specific heat exhibits multiple peaks, which is consistent with a previous study~\cite{FT2,FT3}.
As $|J_2|$ increases, the low-temperature peak shifts to higher temperatures, and the specific heat tends to exhibit a double-peak structure.
At $J_2=0$, the specific heat remains finite even at $T=0.01$.
This suggests that the entropy remains finite, even at low temperatures at approximately $J_2=0$.
Therefore, entropy calculations are also performed.

\begin{figure}[tb]
\includegraphics[width=86mm]{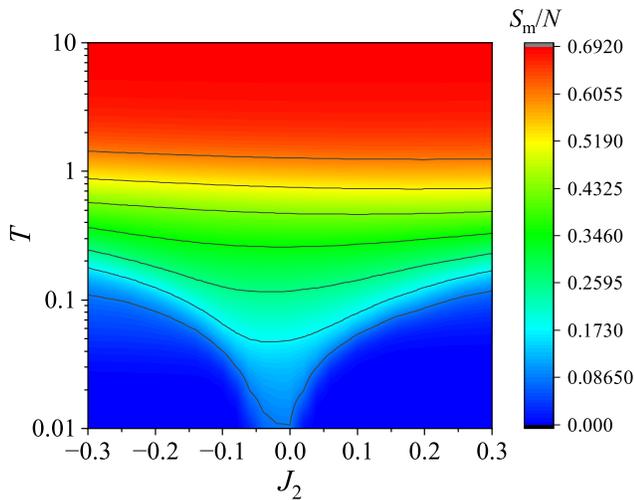}
\caption{
Color plot of the magnetic entropy $S_m$ per site for the $J_1-J_2$ kagome lattice with $N=36$.
The calculation of the entropy dependent on $T$ and $J_2$ was performed at 3001 points for $T$ and 31 points for $J_2$.
\label{Sm}}
\end{figure}

The results of entropy calculations are shown in Fig.~\ref{Sm}.
The entropy remains finite, even at low temperatures for $-0.05 < J_2 < 0.01$.
Beyond this region, the temperature increases rapidly under conditions of low and constant entropy.
These results suggest that the QSL state is realized in the region of $-0.05 < J_2 < 0.01$.
In a classical spin system, at $J_2 > 0$, the ground state is the $\bf Q=0$, $120^\circ$ structure, whereas at $J_2 < 0$, it is the $\sqrt{3} \times \sqrt{3}$ $120^\circ$ structure.
Furthermore, at $J_2=0$, the ground state is infinitely degenerate.
Hence, even in the quantum spin system, the entropy remains finite even at low temperatures at approximately $J_2=0$.

\begin{figure}[tb]
\includegraphics[width=86mm]{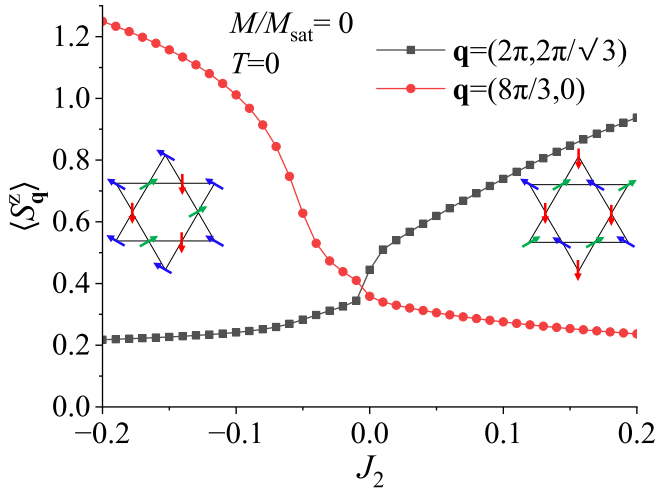}
\caption{
Static spin structure factor $S^z_{\bf q}$ at $T=0$ and $M/M_{sat}=0$ for the $J_1-J_2$ kagome lattice with $N=36$.
The figures on the left and right show the schematics of the $\sqrt{3} \times \sqrt{3}$ $120^\circ$ and $\bf Q=0$, $120^\circ$ structures, respectively.
\label{0Sq}}
 \end{figure}

To investigate the cause of the changes in the specific heat and entropy depending on $J_2$, we calculate the static spin structure factor $S^z_{\bf q}$ at $T=0$, where $S^z_{\bf q} =  \frac{1}{N}\sum_{j} \sum_{k} \it{e}^{\it{i}{\bf q}\cdot ({\bf r}_j-{ \bf r}_k)}S^z_{{\bf r}_j} S^z_{{\bf r}_k}$  with the position vector $\mathbf{r}_j$ and $\mathbf{r}_k$.
If $\langle S^z_{\bf q} \rangle$ is maximal at ${\bf q}=(2\pi, 2\pi/\sqrt3)$, the ground state is expected to have the $\bf Q=0$  structure, whereas if $\langle S^z_{\bf q} \rangle$ is maximal at ${\bf q}=(8\pi/3, 0)$, the ground state is expected to have the $\sqrt{3} \times \sqrt{3}$ structure.
The calculation results for $\langle S^z_{\bf q} \rangle$ are presented in Fig.~\ref{0Sq}.
For $J_2>0$, $\langle S^z_{\bf q} \rangle$ at ${\bf q}=(2\pi, 2\pi/\sqrt3)$ increases as $J_2$ increases, 
whereas for $J_2 < 0$, $\langle S^z_{\bf q} \rangle$ at  ${\bf q}=(8\pi/3, 0)$ increases as $J_2$ decreases.
Thus, the system is expected to exhibit the $\bf Q=0$, $120^\circ$ order for $J_2>0$ and the $\sqrt{3} \times \sqrt{3}$ $120^\circ$ order for $J_2<0$.
This finding is consistent with previous studies \cite{J12k1,J12k2,J12k3,J12k4,J12k5}.
Schematic views of these magnetic structures are presented in Fig.~\ref{0Sq}.
At approximately $J_2=0$, both values are similarly small, which suggests the existence of a QSL state.
These results are consistent with the behavior in which the entropy remains finite even at low temperatures for $-0.05 < J_2 < 0.01$, as shown in Fig.~\ref{Sm}.

 \subsection{Magnetization curve and 1/3 plateau}
\begin{figure}[tb]
\includegraphics[width=80mm]{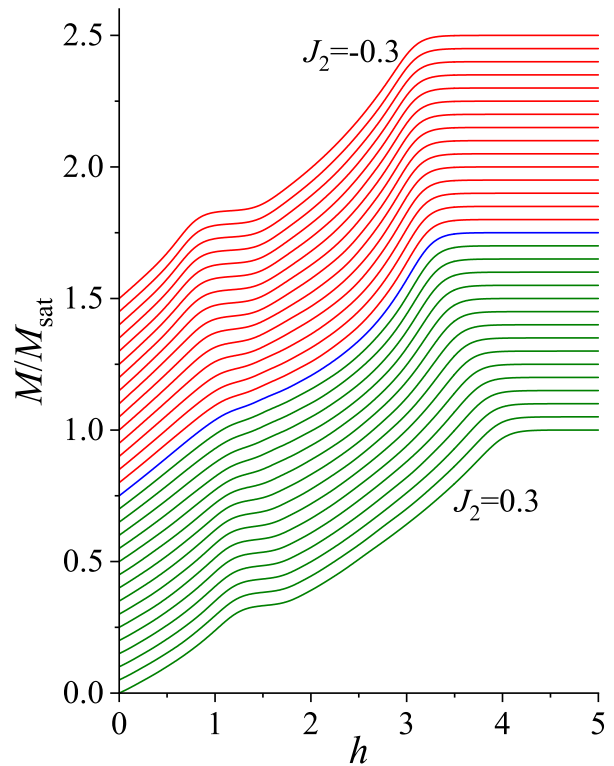}
\caption{
Magnetization curves of the $J_1-J_2$ kagome lattice with $N=36$ for $-0.3 \le J_2 \le 0.3$ at $T=0.1$.
The origin of the vertical axis is shifted by 0.05 for every 0.02 decrease in $J_2$ for better visibility.
The green lines present the results for $J_2 > 0$, the blue line for $J_2=0$, and the red lines for $J_2<0$.
\label{AM-H}}
\end{figure}

 Figure~\ref{AM-H} shows the magnetization curves at $T=0.1$ for $-0.3 \le J_2 \le 0.3$. 
These results exhibit almost no finite size effects as shown in Fig.~\ref{FSE}.
 No apparent plateau is observed around $J_2=0$. 
 As $|J_2|$ increases, an evident 1/3 plateau began to appear. 
This indicates that the magnetic properties change around $J_2=0$, which is similar to the behavior at $h=0$. 
 In a classical spin system, even in a magnetic field, the ground state is the $\bf Q=0$ structure for $J_2 > 0$, and the $\sqrt{3} \times \sqrt{3}$ structure for $J_2 < 0$. 
At $J_2=0$, the ground state is infinitely degenerate. 
In the quantum spin system, a decrease in the degeneracy leads to the stabilization of the 1/3 plateau, as shown in Fig.~\ref{AM-H}.
A 1/3 plateau also appears in the triangular lattice, which is stabilized by the uud structure arising from quantum fluctuations and frustration effects. In contrast, in the $J_1-J_2$ kagome lattice, the 1/3 plateau became unstable when the ground state in the classical limit is infinitely degenerate, that is, at $J_2=0$.
Thus, if an evident 1/3 plateau is not observed in the experimental measurements of spin-1/2 $J_1-J_2$ kagome compounds with antiferromagnetic $J_1$, it can be expected that the magnitude of $J_2$ in that compound is small.

\begin{figure}[tb]
\includegraphics[width=72mm]{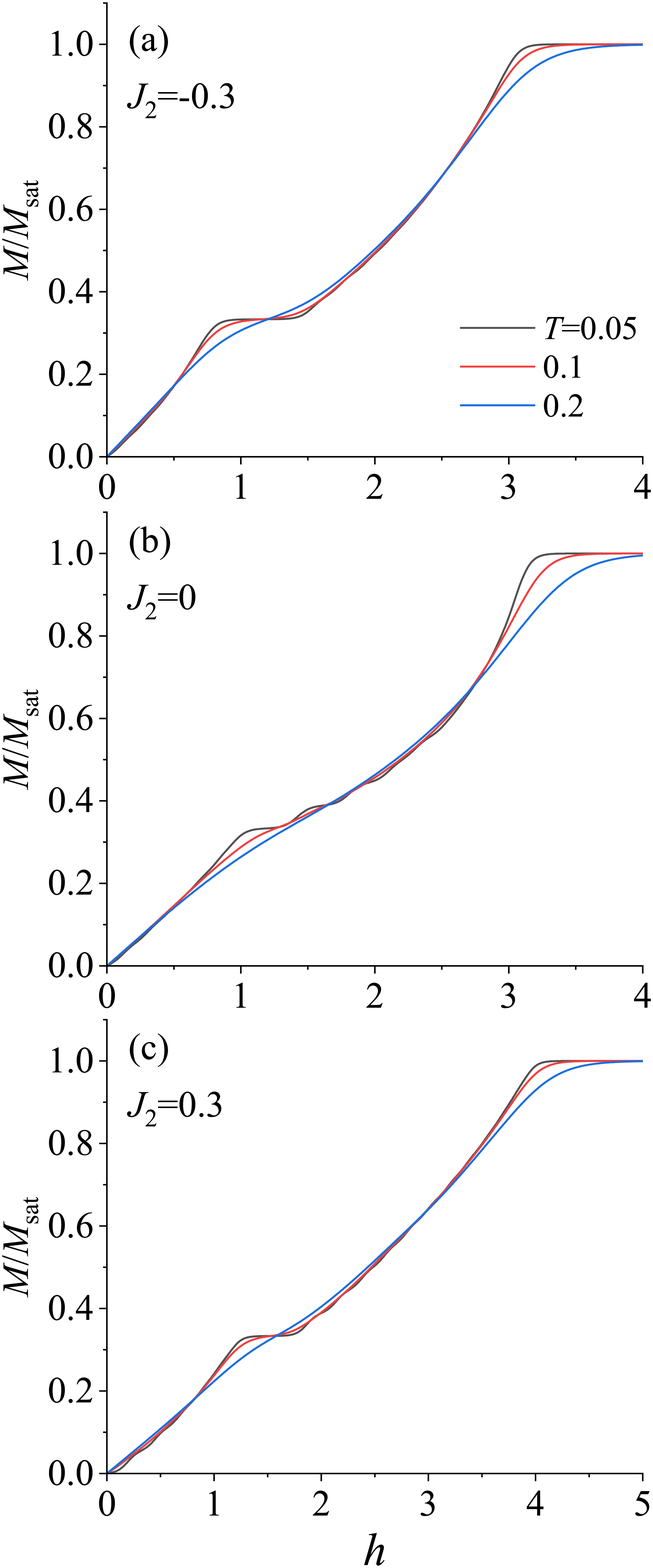}
\caption{
Finite-temperature magnetization curves of the $J_1-J_2$ kagome lattice with $N=36$.
\label{M-H1}}
\end{figure} 

\begin{figure}[tb]
\includegraphics[width=76mm]{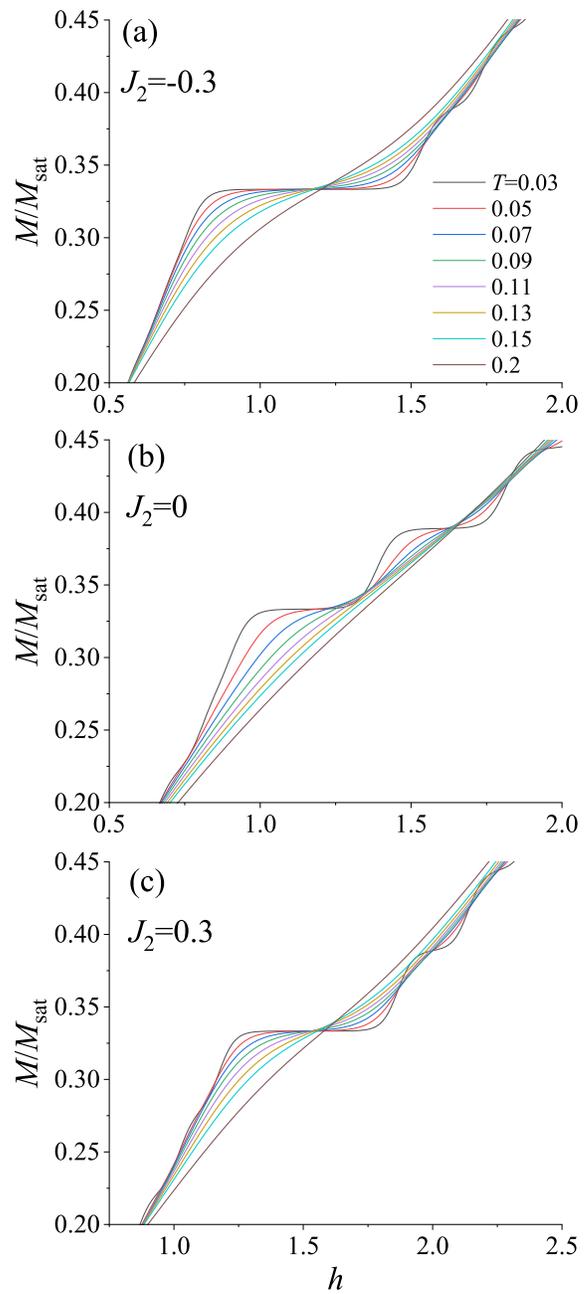}
\caption{
Finite-temperature magnetization curves of the $J_1-J_2$ kagome lattice with $N=36$ around $M/M_{\rm sat}=1/3$.
\label{M-H2}}
\end{figure} 

To investigate the stabilization of the 1/3 magnetization plateau and the asymmetric melting phenomenon in the $J_1-J_2$ kagome lattice, we calculate the magnetization curves at various temperatures for $J_2=-0.3, 0$, and 0.3.
Figure~\ref{M-H1}(a) shows the calculated results for the magnetization curves at $J_2 = -0.3$; \ref{M-H1}(b), at $J_2 = 0$; and \ref{M-H1}(c), at $J_2 = 0.3$.
A 1/3 plateau can be observed at $J_2=0.3$ and $J_2=-0.3$; however, there is no apparent 1/3 plateau at $J_2=0$.
The calculation results down to $T=0.05$ do not reveal the presence of plateaus other than the 1/3 plateau. Further calculations at lower temperatures are required to confirm the presence of the other plateaus.
 However, owing to finite-size effects, the magnetization curve at the thermodynamic limit remains unclear at lower temperatures.

 Figures~\ref{M-H2}(a), \ref{M-H2}(b), and \ref{M-H2}(c) show enlarged views of the finite-temperature magnetization curves around $M/M_{\rm sat}=1/3$.
At $J_2 = \pm0.3$, the 1/3 plateau gradually melts in a nearly symmetric manner as the temperature increases.
However, at $J_2 = 0$, the deviation of the magnetization from 1/3 starts on the left side of the 1/3 plateau, which indicates an asymmetric melting behavior.
This result is consistent with previous studies~\cite{FT3,FT4,FT5}.
Hence, it is evident that the phenomenon of asymmetric melting occurs only around $J_2=0$.
Instead of a plateau, a ramp appears at  $M/M_{\rm sat}=1/3$ and thus the slope is different to the left and right. Consequently, one anticipates a difference in the density of states on each side. While in the cases where a plateau exists, the slope remains relatively consistent at both ends of the plateau, implying a comparable density of states on either side. This conjecture is later confirmed through numerical calculations.

\begin{figure}[tb]
\includegraphics[width=86mm]{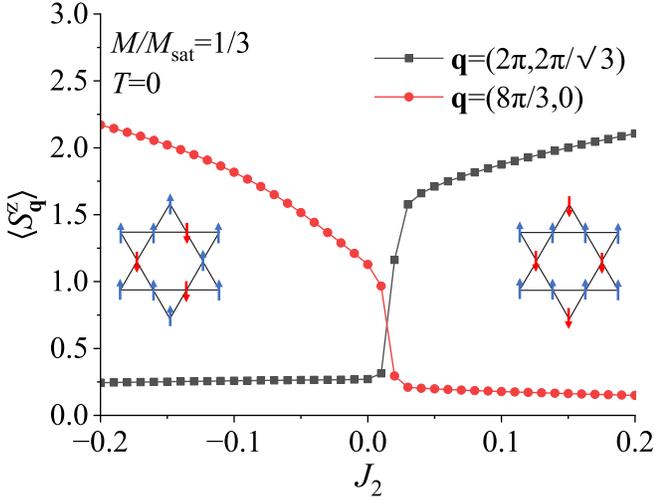}
\caption{
Static spin structure factor $S^z_{\bf q}$ at $T=0$ and $M/M_{sat}=1/3$ for the $J_1-J_2$ kagome lattice with $N=36$.
The figures on the left and right show the schematics of the $\sqrt{3} \times \sqrt{3}$ uud and $\bf Q=0$ uud structures, respectively.
\label{1-3Sq}}
\end{figure}

To investigate the singularity around $J_2=0$, we examine the magnetic structure of the 1/3 plateau at $T = 0$ by calculating $\langle S^z_{\bf q} \rangle$ at $M/M_{\rm sat}=1/3$.
The calculation results for $\langle S^z_{\bf q} \rangle$ are presented in Fig.~\ref{1-3Sq}. 
For $J_2 > 0$, as $J_2$ increases the $\bf Q=0$ uud structure stabilizes, whereas for $J_2 < 0$, as 
$|J_2|$ increases, the $\sqrt{3} \times \sqrt{3}$ uud structure stabilizes. 
Schematics of these magnetic structures are shown in Fig.~\ref{1-3Sq}.
Both these structures are semiclassical magnetic structures with uud structures similar to the 1/3 plateau of the triangular lattice. 
Therefore, when $|J_2|$ is large, the 1/3 plateau melts symmetrically, similar to the triangular lattice.
In Fig.~\ref{1-3Sq}, a phase transition occurs at $J_2\sim0.015$. 
The 1/3 plateau becomes unstable at approximately $J_2=0$ because of its proximity to the phase transition point.

At $J_2 = 0$, $\langle S^z_{\bf q} \rangle$ reaches  its highest value at  ${\bf q}=(8\pi/3, 0)$. This result does not negate previous studies that the 1/3 plateau exhibits a VBC structure~\cite{KH3,KH4}.
Although this study does not provide conclusive evidence, we anticipate a phase transition between the $\sqrt{3} \times \sqrt{3}$ uud phase and the VBC phase to occur at a certain value of $J_2$ in the thermodynamic limit, because the ground state is expected to be the $\sqrt{3} \times \sqrt{3}$ uud state when $J_2$ is sufficiently negative.

\begin{figure}[tb]
\includegraphics[width=86mm]{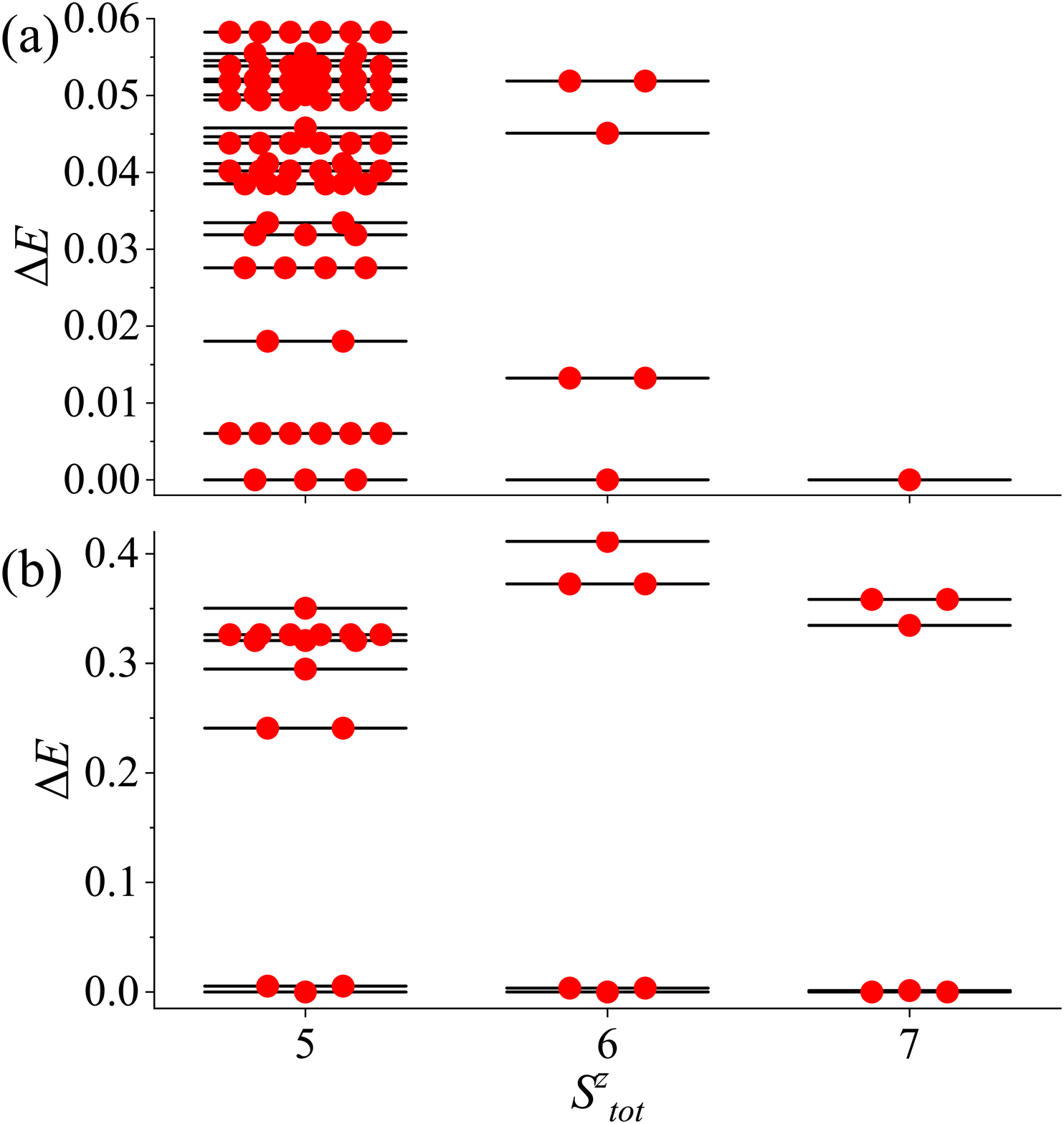}
\caption{
Low-energy excitation spectra of the $J_1-J_2$ kagome lattice with $N=36$ for the states with $S^z_{tot} = 5, 6,$ and 7 at $J_2=0$ (a) and  $J_2 = -0.3$ (b).
The horizontal bars indicate the energy gap $\Delta E$.
The number of filled red circles represents the degeneracy.
Note that the vertical scales are different between (a) and (b).
\label{Es}}
\end{figure}

We calculate low-energy excitation spectra to investigate further the reasons for the asymmetric melting of the 1/3 plateau at $J_2=0$ and the symmetric melting at $J_2=\pm0.3$ as shown in Fig.~\ref{M-H2}.
Figure~\ref{Es} shows the low-energy excitation spectra for states with $S^z_{tot} = 5, 6,$ and 7 at $J_2=0$ and $J_2 = -0.3$. Here, $S^z_{tot} = 6$ corresponds to $M/M_{sat} = 1/3$.
At $J_2=0$, for $S^z_{tot} = 7$, all the excitation energies are $\Delta E > 0.1$, whereas for $S^z_{tot} = 5$, there are 76 states in $\Delta E \le 0.06$. 
This indicates that the states with $S^z_{tot} = 5$ are entropically favored over those with $S^z_{tot} = 7$.
Therefore, these energy spectra lead to the asymmetric melting of the 1/3 plateau.
In contrast, at $J_2 = -0.3$, similar energy spectra are observed regardless of $S^z_{tot} = 5, 6,$ and 7.
 They all exhibit pseudo-triple degeneracy corresponding to the $\sqrt{3} \times \sqrt{3}$ structure, and the excitation energies are $\Delta E > 0.2$. 
 These energy spectra are different from those at $J_2 = 0$. 
 Hence, the 1/3 plateau at $J_2 = -0.3$ melts symmetrically. 

\section{Summary}
\label{sec5}
Inspired by recent studies on the finite-temperature properties of the kagome lattice, we investigated the finite-temperature properties of the spin-1/2 $J_1-J_2$ kagome lattice using OFTLM.
At $J_2 = 0$, the specific heat exhibited a multipeak structure, but as $|J_2|$ increased, 
the multipeak structure transitioned into a double-peak structure.
For $-0.05 < J_2 < 0.01$, the magnetic entropy exhibited a finite value even at low temperatures, indicating the presence of a QSL.
This range of $-0.05<J_2<0.01$ is slightly narrower than the range of $-0.1\lesssim J_2 \lesssim 0.1$ in the previous studies~\cite{J12k2,J12k3,J12k4,J12k5}.
In the magnetization curve, we observed an asymmetric melting phenomenon with a 1/3 plateau around $J_2 = 0$. In contrast, as $|J_2|$ increased, the 1/3 plateau melted symmetrically, exhibiting apparent flatness.
In the 1/3 plateau, the $\bf Q=0$ uud state was stabilized for $J_2 > 0$, whereas the $\sqrt{3} \times \sqrt{3}$ uud state was stabilized for $J_2 < 0$.
Contrary to the conventional understanding that as the degeneracy of the classical ground state increases, quantum effects become more pronounced, leading to the emergence of a magnetization plateau, our findings demonstrated that the 1/3 plateau stabilizes as the degeneracy is reduced.
In the future, by comparing our results with experimental data, we will be able to determine the magnitude of $J_2$ in spin-1/2 $J_1-J_2$ kagome compounds with antiferromagnetic $J_1$. 
We believe that our results contribute to a deeper understanding of the physics of kagome lattices.

\begin{acknowledgments}
We thank Y. Ishii, H. Yoshida, and Y. Fukumoto for useful discussions.
We also thank the Supercomputer Center, the Institute for Solid State Physics, the University of Tokyo for the use of the facilities.
\end{acknowledgments}


\begin{thebibliography}{99}
\addcontentsline{toc}{section}{References}
\bibitem{KLP1}  P. Lecheminant, B. Bernu, C. Lhuillier, L. Pierre, and P. Sindzingre, Phys. Rev. B {\bf 56}, 2521 (1997).
\bibitem{KLP2}
C. Waldtmann, H.-U. Everts, B. Bernu, C. Lhuillier, P. Sindzingre, P. Lechminant, and L. Pierre, 
Eur. Phys. J. B {\bf 2} 501 (1998).
\bibitem{KLP3}  L. Balents, Nature {\bf 464}, 199 (2010).


\bibitem{Z2-1} S. Yan, D. A. Huse, and S. R. White, Science {\bf 332}, 1173 (2011). 
\bibitem{Z2-2} S. Depenbrock, I. P. McCulloch, and U. Schollwock, 
Phys. Rev. Lett. {\bf 109}, 067201 (2012). 
\bibitem{U1-1} Y. Ran, M. Hermele, P. A. Lee, and X. G. Wen, Phys. Rev. Lett. {\bf 98}, 117205 (2007). 
\bibitem{U1-2} Y. Iqbal, F. Becca, and D. Poilblanc, Phys. Rev. B {\bf 83}, 100404(R) (2011).
\bibitem{U1-3} Y. Iqbal, F. Becca, S. Sorella, and D. Poilblanc, Phys. Rev. B {\bf 87}, 060405(R) (2013). 
\bibitem{U1-4} Y.-C. He, M. P. Zaletel, M. Oshikawa, and F. Pollmann, Phys. Rev. X {\bf 7}, 031020 (2017).
\bibitem{VBC-1} J. B. Marston and C. Zeng, J. Appl. Phys. {\bf 69}, 5962 (1991). 
\bibitem{VBC-2} R. R. P. Singh and D. A. Huse, Phys. Rev. B {\bf 76}, 180407(R) (2007).
\bibitem{VBC-3} K. Hwang, Y. B. Kim, J. Yu, and K. Park, Phys. Rev. B {\bf 84}, 205133 (2011). 
%
\bibitem{KH1} A. Honecker, J. Schulenburg, and J. Richter, J. Phys.: Condens. Matter {\bf16} and S749 (2004). 
\bibitem{KH2} H. Nakano and T. Sakai, J. Phys. Soc. Jpn. {\bf 79}, 053707 (2010). 
\bibitem{KH3} S. Capponi, O. Derzhko, A. Honecker, A. M. Lauchli, and J. Richter, Phys. Rev. B {\bf 88}, 144416 (2013).
\bibitem{KH4} S. Nishimoto, N. Shibata, and C. Hotta, Nat. Commun. {\bf 4}, 2287 (2013).
\bibitem{KH5} T. Picot, M. Ziegler, R. Orus, and D. Poilblanc, Phys. Rev. B {\bf 93}, 060407(R) (2016).
\bibitem{KH6} H. Nakano and T. Sakai, J. Phys. Soc. Jpn. 87, 063706 (2018).
%
\bibitem{FT1}  N. Elstner and A. P. Young, Phys. Rev. B {\bf 50}, 6871 (1994).
\bibitem{FT2}  T. Shimokawa and H. Kawamura, J. Phys. Soc. Jpn. {\bf 85}, 113702 (2016).
\bibitem{FT3} J. Schnack, J. Schulenburg, and J. Richter, Phys. Rev. B {\bf 98}, 094423 (2018).
\bibitem{FT4} T. Misawa, Y. Motoyama, and Y. Yamaji, Phys. Rev. B {\bf 102}, 094419 (2020).
\bibitem{FT5} H. Schluter, J. Richter, and J. Schnack, J. Phys. Soc. Jpn. {\bf 91}, 094711 (2022).
%
\bibitem{ke1} M. P. Shores, E. A. Nytko, B. M. Bartlett, and D. G. Nocera, J. Am. Chem. Soc. {\bf 127}, 13462 (2005).
\bibitem{ke2} T.-H. Han, J. S. Helton, S. Chu, D. G. Nocera, J. A. Rodriguez-Rivera, C. Broholm, and Y. S. Lee, Nature {\bf 492} 406 (2012).
%
\bibitem{ke3} K. Morita, M. Yano, T. Ono, H. Tanaka, K. Fujii, H. Uekusa, Y. Narumi, and K. Kindo, J. Phys. Soc. Jpn. {\bf 77}, 043707 (2008).
\bibitem{ke4} T. Ono, K. Morita, M. Yano, H. Tanaka, K. Fujii, H. Uekusa, Y. Narumi, and K. Kindo, Phys. Rev. B {\bf 79}, 174407 (2009).
\bibitem{ke5} K. Matan, T. Ono, G. Gitgeatpong, K. de Roos, P. Miao, S. Torii, T. Kamiyama, A. Miyata, A. Matsuo, K. Kindo, S. Takeyama, Y. Nambu, P. Piyawongwatthana, T. J. Sato, and H. Tanaka, Phys. Rev. B {\bf 99}, 224404 (2019).
\bibitem{ke6} M. Goto, H. Ueda, C. Michioka, A. Matsuo, K. Kindo, and K. Yoshimura, Phys. Rev. B {\bf 94}, 104432 (2016).
\bibitem{ke7} R. Shirakami, H. Ueda, H. O. Jeschke, H. Nakano, S. Kobayashi, A. Matsuo, T. Sakai, N. Katayama, H. Sawa, K. Kindo, C. Michioka, and K. Yoshimura,
Phys. Rev. B {\bf 100}, 174401 (2019).
%
\bibitem{ke8} W. Sun, Y.-X. Huang, S. Nokhrin, Y. Pan, and J.-X. Mi, J. Mater. Chem. C {\bf 4}, 8772 (2016).
%
\bibitem{ke9} R. Okuma, T. Yajima, D. Nishio-Hamane, T. Okubo, and Z. Hiroi, Phys. Rev. B {\bf 95}, 094427 (2017).
%
\bibitem{ke10} H. Yoshida, N. Noguchi, Y. Matsushita, Y. Ishii, Y. Ihara, M. Oda, H. Okabe, S. Yamashita, Y. Nakazawa, A. Takata, T. Kida, Y. Narumi, and M. Hagiwara, J. Phys. Soc. Jpn. {\bf 86}, 033704 (2017).
\bibitem{ke11} R. Okuma, D. Nakamura, T. Okubo, A. Miyake, A. Matsuo, K. Kindo, M. Tokunaga, N. Kawashima, S. Takeyama, and Z. Hiroi, Nat. Commun.  {\bf 10}, 1229 (2019).
\bibitem{ke12} L. M. Volkova and D. V. Marinin, J. Phys.: Condens. Matter {\bf 33} 415801 (2021).
\bibitem{ke13} H. K. Yoshida, J. Phys. Soc. Jpn. {\bf 91}, 101003 (2022).

\bibitem{J12k1} R. Suttner, C. Platt, J. Reuther, and R. Thomale, Phys. Rev. B {\bf 89}, 020408(R) (2014). 
\bibitem{J12k2} S.-S. Gong, W. Zhu, L. Balents, and D. N. Sheng, Phys. Rev. B {\bf 91}, 075112 (2015).
\bibitem{J12k3} F. Kolley, S. Depenbrock, I. P. McCulloch, U. Schollwock, and V. Alba, Phys. Rev. B {\bf 91}, 104418 (2015).
\bibitem{J12k4} P. Prelovsek, K. Morita, T. Tohyama, and J. Herbrych, Phys. Research {\bf 2}, 023024 (2020).
\bibitem{J12k5} Y. Iqbal, F. Ferrari, A. Chauhan, A. Parola, D. Poilblanc, and F. Becca, Phys. Rev. B {\bf 104}, 144406 (2021).

\bibitem{OFTL1} K. Morita and T. Tohyama, Phys. Research  {\bf 2}, 013205 (2020).
\bibitem{OFTL2} K. Morita, Phys. Rev. B  {\bf 105}, 064428 (2022).


\bibitem{ftla1} Y. Shibata, T. Tohyama, and S. Maekawa, Phys. Rev. B {\bf 64}, 054519 (2001).
\bibitem{ftla2} N. Shannon, B. Schmidt, K. Penc, and P. Thalmeier, Eur. Phys. J. B {\bf 38}, 599 (2004).
\bibitem{ftla3} I. Zerec, B. Schmidt, and P. Thalmeier, Phys. Rev. B {\bf 73}, 245108 (2006).
\bibitem{ftla4} B. Schmidt, P. Thalmeier, and N. Shannon, Phys. Rev. B {\bf 76}, 125113 (2007).
\bibitem{ftla5} J. Schnack and O. Wendland, Eur. Phys. J. B {\bf 78}, 535 (2010).
\bibitem{ftla6} J. Schnack and C. Heesing, Eur. Phys. J. B {\bf 86}, 46 (2013).
\bibitem{ftla7} O. Hanebaum and J. Schnack, Eur. Phys. J. B {\bf 87}, 194 (2014).
\bibitem{ftla8} T. Munehisa, World J. Condens. Matter Phys. {\bf 7}, 11 (2017). 
\bibitem{ftla9} J. Schnack, J. Richter, and R. Steinigeweg, Phys. Research {\bf 2}, 013186 (2020).
\bibitem{ftla10} K. Seki and S. Yunoki, Phys. Rev. B {\bf 101}, 235115 (2020).
\bibitem{ftla11} J. Schnack, J. Schulenburg, A. Honecker, and J. Richter, Phys. Rev. Lett. {\bf 125}, 117207 (2020).
\bibitem{ftla12} J. Richter, O. Derzhko, and J. Schnack, Phys. Rev. B {\bf 105}, 144427 (2022).
\bibitem{ftla13} J. Richter and J. Schnack, Phys. Rev. B {\bf 107}, 245115 (2023).
























\end{thebibliography}
\end{document}